\documentclass[12pt]{article}

\usepackage{epsfig}

\usepackage{amssymb}
\usepackage{amsfonts}

\usepackage{color}
\def\be{\begin{equation}}
\def\ee{\end{equation}}
\def\ba{\begin{array}{c}}
\def\ea{\end{array}}

\newcommand{\bea}{\begin{eqnarray}}
\newcommand{\eea}{\end{eqnarray}}

\newcommand{\bbr}{\br\!\br}
\newcommand{\kkt}{\kt\!\kt}

\newcommand{\pkt}{\!\!\succ\,\,}
\newcommand{\kt}{\rangle}
\newcommand{\br}{\langle}

\begin{document}

\begin{center}

{\Large \bf

Three
alternative model-building strategies using
 quasi-Hermitian time-dependent observables

}

\end{center}

\vspace{0.8cm}

\begin{center}

  {\bf Miloslav Znojil}$^{a,b}$

\end{center}

 $^{a}$  {Department of Physics, Faculty of
Science, University of Hradec Kr\'{a}lov\'{e}, Rokitansk\'{e}ho 62,
50003 Hradec Kr\'{a}lov\'{e},
 Czech Republic}

 $^{b}$
{The Czech Academy of Sciences,
 Nuclear Physics Institute,
 Hlavn\'{\i} 130,
250 68 \v{R}e\v{z}, Czech Republic,
{e-mail znojil@ujf.cas.cz}

\vspace{10mm}

%\newpage

%\textcolor{black}{...}

\subsection*{Abstract}

In the conventional
(so called
Schr\"{o}dinger-picture)
formulation of quantum theory
the operators of observables
are chosen
self-adjoint and time-independent.
In the recent
innovation of the theory
the operators
can be not only
non-Hermitian but also time-dependent.
The formalism
(called non-Hermitian interaction-picture, NIP)
requires a separate description of the evolution of the
time-dependent states $\psi(t)$
(using
Schr\"{o}dinger-type equations)
as well as of the time-dependent observables $\Lambda_j(t)$, $j=1,2,\ldots,K$
(using Heisenberg-type equations).
In the unitary-evolution dynamical regime of our interest,
both of the respective generators of the evolution
(viz., in our notation, the Schr\"{o}dingerian generator $G(t)$
and the Heisenbergian generator $\Sigma(t)$) have,
in general, complex spectra. Only the spectrum of their superposition
remains real. Thus, only the
observable superposition $H(t)=G(t)+\Sigma(t)$
(representing the
instantaneous energies)
should be called Hamiltonian.
In applications, nevertheless,
the mathematically consistent models
can be based not only on the initial knowledge of the energy operator $H(t)$
(forming a ``dynamical'' model-building strategy)
but also, alternatively, on the knowledge of the Coriolis force $\Sigma(t)$
(forming a ``kinematical'' model-building strategy),
or on the initial knowledge of the Schr\"{o}dingerian generator $G(t)$
(forming, for some reasons, one of the most
popular strategies in the literature).
In our present paper
every such a choice
(marked as ``one'', ``two'' or ``three'', respectively)
is shown to
lead to a construction recipe
with a specific range of applicability.

\subsection*{Keywords}

quantum theory of unitary systems;
non-Hermitian interaction representation;
non-stationary physical inner products;
model-building classification;

\newpage

\section{Introduction}

One of the sources of inspiration of our present study was a compact
review \cite{Styer} of the history of alternative formulations of
quantum mechanics. In their paper dated 2002 the authors asked the
question of how many formulations of quantum mechanics do we
have\textcolor{black}{.} For pedagogical reasons, nevertheless,
their list remained incomplete. Surprisingly enough, it did not
include the Dirac's ``intermediate-'' {\it alias\,}
``interaction-picture'' (IP) form of the Hermitian theory. The
authors also avoided any reference to the innovative paper
\cite{Geyer} in which the conventional lists of the available
alternative formulations of quantum mechanics were complemented, as
early as in 1992, by a manifestly non-Hermitian reformulation of
Schr\"{o}dinger picture (NSP, see also its more recent comprehensive
review in \cite{ali}).

The latter omission was in fact not too surprising because the NSP
(also known as ``quasi-Hermitian''  \cite{Geyer,Dieudonne})
formalism has only been developed between the years 1998 and 2007
when Bender with multiple coauthors made the idea widely known and
popular \cite{BB,Carl}. In spite of certain scepticism among
specialists (as verbalized, e.g., by Streater \cite{Streater} or,
more recently, by several mathematicians
\cite{Trefethen,Siegl,Viola,Iveta,Uwe}), Bender with his coauthors
persuaded the quantum-physics community that there exists a broad
class of innovative stationary realizations of quantum theory
(including, importantly, quantum field theory \cite{BM}) in which
the manifestly non-Hermitian candidates $H$ for the Hamiltonians
with real spectra could be phenomenologically appealing as well as
mathematically sufficiently user-friendly (cf. also the newer
reviews of the field in
\textcolor{black}{\cite{Christodoulides,Carlbook,book,ju,judr,judrb,judrc}}).

Before the year 2008 the next-step transition to the non-stationary
non-Hermitian theory
has been considered impossible \cite{PLB}.
At the same time, the idea of the
{\em stationary\,} unitary evolution ``in non-Hermitian disguise''
has been widely accepted.
People realized that
such a formulation of the theory remains
equivalent to its standard textbook predecessors.
For this reason, the presentation of the innovation could have started from the
conventional stationary Schr\"{o}dinger equation
 \be
 {\rm i}\,\frac{d}{dt}\,|\psi^{}(t)\pkt
 =
 \mathfrak{h}\,|\psi^{}(t)\pkt
 \,,
 \ \ \ \ \mathfrak{h}=\mathfrak{h}^\dagger\neq \mathfrak{h}(t)\,.
 \label{pCauchy}
 \ee
The generalization
(attributed, often, to Dyson \cite{Dyson}) proved based just on a
replacement of such an equation by its non-Hermitian (or, better, hiddenly
Hermitian) upgrade mediated by an invertible time-independent
mapping $\Omega\neq \Omega(t)$,
 \be
 {\rm i}\,\frac{d}{dt}\,|\psi^{}(t)\kt
 =
 H\,|\psi^{}(t)\kt
 \,,
 \ \ \ \ H=\Omega^{-1}\,\mathfrak{h}\,\Omega
 \neq H^\dagger\,,\ \ \ H \neq H(t)\,.
 %\ \ \ \ H^\dagger\,\Theta=\Theta\,H\,.
 \label{hCauchy}
 \ee
During the year 2008 the scientific community
became prepared to accept the proposal
of making the non-Hermitian theory {\em non-stationary} \cite{timedep}.
In the recent application of this approach
to the so called wrong-sign interaction potentials \cite{Entropy}
we pointed out
that
within the corresponding
form of quantum theory
called non-Hermitian
interaction picture (NIP, see also its compact review in \cite{NIP}),
the unitarity of the evolution of the so called closed quantum systems
can still be guaranteed in consistent manner.

In contrast to our preceding NIP-based paper \cite{Entropy}, its
present continuation will be example-independent. Only a few remarks
on possible applications will be added, \textcolor{black}{mainly} in
Appendix A. In the main body of our new, more methodically oriented
paper we will discuss the three main model-building strategies. In a
systematic manner the presentation of our results will start in
section~\ref{motivation} in which we will review the basic ideas
behind the existing hiddenly-Hermitian quantum theories. In
sections~\ref{hamb} - \ref{Gpic} we will then outline the three
respective construction options emphasizing, in each of them, the
necessity of a clear separation of what is assumed and postulated
from what is calculated, reconstructed and deduced.

A compact summary of our considerations is finally
added in section \ref{progresivni}.

\section{The abstract NIP quantum theory\label{motivation}}

\subsection{The concept of non-stationary non-Hermitian observables}

In conventional textbooks one often reads about
the choice of a ``picture'' {\it alias\,} ``representation''
of quantum mechanics (cf., e.g.,
\cite{Messiah}). Let us temporarily return, therefore, to the
Hermitian theory. One then usually
mentions just the Schr\"{o}dinger picture (SP)
and the Heisenberg picture (HP). Sometimes, another
option is presented
under the name of ``intermediate picture'' (IP, cf. pp.
321 - 322 in \cite{Messiah}). In this case
one is assumed to split a given self-adjoint
Hamiltonian in its two separate self-adjoint components,
$\mathfrak{h}=\mathfrak{h}_S+\mathfrak{h}_H$. Typically,
$\mathfrak{h}_S$ is designed to control the evolution of states
(i.e., it appears
as a generator in a Schr\"{o}dinger-type equation)
while $\mathfrak{h}_H$ is interpreted as entering the
Heisenberg-type equations for
the relevant and, necessarily, time-dependent but still self-adjoint
observables~\cite{Messiah}.

After one moves to the above-mentioned non-Hermitian
(or, using a mathematically more precise
terminology, quasi-Hermitian \cite{Geyer,Dieudonne})
reformulations of quantum theory,
a part of the terminology survives.
In particular, in the non-Hermitian interaction picture (NIP, \cite{NIP})
we still encounter
the
Heisenberg's generator (say, $\Sigma(t)$, i.e., the operator
controlling the time-evolution of the non-stationary
NIP observables \cite{Bishop}) as well as the
Schr\"{o}dinger's generator (to be denoted as $G(t)$)
entering the Schr\"{o}dinger-type evolution equations.

The new feature \textcolor{black}{of} the more general non-Hermitian
theory is that one can speak about a quantum Coriolis force
$\Sigma(t)$ \cite{Coriolis} filling the interval or space between
its NSP extreme $\Sigma^{(NSP)}(t)=0$ and its non-Hermitian HP
extreme such that $\Sigma^{(NHP)}(t)=H^{(NHP)}(t)$
\cite{NHeisenberg}. Moreover, one can formally define, not quite
expectedly \cite{SIGMA}, the superposition of the generators
 \be
 H(t)=G(t)+\Sigma(t)\,.
 \label{decoim}
 \ee
This operator
carries a clear physical meaning
of an isospectral avatar of its self-adjoint SP partner Hamiltonian
$\mathfrak{h}$ (cf. Eq.~(\ref{hCauchy})).
The relevance of such a property has been emphasized
in \cite{Entropy} where we paid attention to the very specific
non-Hermitian anharmonic-oscillator models.
We arrived there at the
conclusion that one cannot easily transfer the ``picture-selection''
experience gained during the study of the special ``wrong-sign''
oscillators to the other non-stationary quasi-Hermitian quantum
systems. Every NIP-described unitary quantum system has to be
treated as specific.

In both of the above-mentioned NSP and NHP special cases
relation (\ref{decoim}) degenerates to
an identity. We either get the coincidence
$G^{(NSP)}(t)\equiv H^{(NSP)}(t)$ with disappearing $\Sigma^{(NSP)}(t)=0$,
or $\Sigma^{(NHP)}(t)\equiv H^{(NHP)}(t)$ with disappearing
$G^{(NHP)}(t)=0$.
Incidentally, it is worth adding that
the widely used attribute ``non-Hermitian'' of the
theory can be misleading because we mean non-Hermitian in our
working space ${\cal H}_{(unphysical)}$ but not in the correct space
of states ${\cal H}_{(physical)}$.
A better name would certainly be ``hiddenly Hermitian'' theory,
meaning that
the operator
$\Theta=\Omega^\dagger\Omega$
of the inner-product metric in ${\cal H}_{(physical)}$
is
nontrivial, $\Theta \neq I$ \cite{Geyer,ali}.

In the most general NIP setting the latter operator is also
assumed manifestly time-dependent,
$\Theta=\Theta(t)$.
The flexibility is enhanced because $G^{(NIP)}(t)\neq 0 \neq \Sigma^{(NIP)}(t)$.
At the same time,
even in the NIP framework
one can distinguish
between the different ways of encoding
the input information about dynamics into operators.
Thus, in our present paper we will speak
about a ``strategy number one''
(cf. section \ref{hamb} below), a ``strategy number two''
(cf. section \ref{Corio}) and ``strategy number three''
(discussed in section \ref{Gpic}).
In other words, we are now going to propose that every choice of {\em one\,}
of the operators in Eq.~(\ref{decoim}) might be interpreted as leading to
a specific eligible representation of the (by assumption, unitary) evolution.

Our forthcoming analysis of a triplet of refined NIP formulations of quantum
theory was in fact motivated by the recent growth
of the diversity of applications
of the  non-stationary versions of the
non-Hermitian
operators in the various branches of physics \cite{book}.
In these applications the building of models appeared often
restricted by a requirement of having an exact, non-numerical form
of the model.
Such a solvability requirement made the conclusions
rather special and model-dependent.
In what follows we will accept, therefore, a different attitude.
We will try to separate, clearly, the form and extent of the
input information about the system
from a systematic and consistent step-by-step reconstruction of the
consequences of the assumptions.

As we already indicated, we will arrive at
three alternative model-building strategies.
In the construction
strategy number one as described in section \ref{hamb}
we will accept the most traditional ``dynamical''
point of view of Scholtz et al \cite{Geyer}.
We will emphasize that, in some sense,
the
observability property makes
the energy-representing operator
$H^{(NIP)}(t)=H^{(NIP)}_{(one)}(t)$ a
unique candidate for being called Hamiltonian.
Indeed, its time-dependent choice determines the
quantum system's dynamics in a phenomenologically satisfactory manner
even in the non-stationary scenario.

In the alternative construction
strategy number two as presented in section \ref{Corio}
we will start from the knowledge of the
physical-Hilbert-space ``kinematics''. Having the knowledge of the
Coriolis-force operator
$\Sigma^{(NIP)}(t)=\Sigma^{(NIP)}_{(two)}(t)$
at all times at our disposal
we will
reconstruct the eligible forms of the dynamics
in a way which will be shown to be
exceptionally straightforward.

In the
strategy number three
we will assume that a key technical
as well as phenomenological role is played by the
time-dependence of the states.
Thus, what is assumed to be
given in advance is the  Schr\"{o}dingerian generator
$G^{(NIP)}(t)=G^{(NIP)}_{(three)}(t)$.

\subsection{The physical inner-product metric }

%{The guarantees of the unitarity of the evolution}

In the NIP picture, both the states $\psi$ (i.e., the elements of a
suitable Hilbert space) and the operators
$\Lambda=\Lambda_j$ (representing observables) are allowed to vary
with time. The respective generators of evolution, i.e., an operator
$G(t)$ entering the Schr\"{o}dinger-type equation for
$\psi=\psi(t)$, and another operator $\Sigma(t)$ in the
Heisenberg-type equation for $\Lambda_j=\Lambda_j(t)$ may be (and,
in the literature, quite often are) both called ``Hamiltonians''.
For this reason it may be useful to try to avoid misunderstandings
by speaking, more explicitly, about a ``non-Hermitian
time-dependent ``Schr\"{o}dinger-equation Hamiltonian''
$G(t)=G^{(NIP)}(t)$ in the former case. We may also need to
amend the denotation of
the ``Heisenberg-equation-Hamiltonian'',  $\Sigma(t) =\Sigma^{(NIP)}(t)$.

In the unitary evolution scenario
both of the auxiliary NIP generators $G(t)$
and $\Sigma(t)$ are just auxiliary
and non-observable. Their spectra may be
complex -- for illustration see, e.g., the schematic examples in
\cite{2by2,3by3}. Only
their instantaneous-energy-representing sum (\ref{decoim})
may be considered, in a way shown in \cite{timedep,SIGMA}, observable.
For this reason we will
call such an operator ``observable
Hamiltonian'' or simply ``Hamiltonian''.

We will
interpret the relationship~(\ref{decoim}) between the three eligible
operators (i.e., between the ``Hamiltonians'' in a broader sense)
as a starting point of the theoretical constructive efforts.
We propose that in such a setting one
picks up simply one of these operators as ``known'',
i.e., as an operator carrying
a decisive portion of the ``physical'' input
information about the unitary quantum
system in question.
In this manner one arrives at the three alternative
quantum-model-building strategies
as described below.

As long as the form of the NIP Hamiltonian $H(t)$ of
Eq.~(\ref{decoim}) has to be flexible and, in particular, not
necessarily Hermitian, the underlying
``working'' Hilbert space (say, ${\cal F}$)
can be declared, in general, unphysical,
playing just the role of a mathematical
tool,
${\cal F} \equiv {\cal H}_{(unphysical)}$.
In 1998, such an idea of working with
observables in a mathematically friendlier
non-Hermitian representation has been made particularly
attractive and popular by Bender with Boettcher \cite{BB}.
One of its main consequences is that our auxiliary,
computation-friendly Hilbert space ${\cal F}$ (i.e., in many
realistic models, just $L^2(\mathbb{R}^d)$
\cite{Christodoulides,Carlbook}) has to be complemented by
another, correct
physical alternative $ {\cal H}_{(physical)}\equiv
{\cal H}\neq {\cal F}$.
In the words of the
older comprehensive review \cite{Geyer} one only has to clarify
the relationship between ${\cal F}$ and ${\cal H}$ by
establishing ``a criterion for a set of non-Hermitian operators''
(i.e., for the above-mentioned set of operators $\Lambda_j$) ``to
constitute a consistent quantum mechanical system'' which ``involves
the construction of a [physical Hilbert-space] metric'',
i.e., which involves a
representation of ${\cal H}$ in ${\cal F}$.
In our present notation
this means that there must exist a suitable
``inner-product-metric'' operator $\Theta$ such
that
 \be
 \Lambda_j^\dagger\,\Theta =\Theta\,\Lambda_j\,,\ \ \ \ j = 0, 1,
 \ldots, J\,.
 \label{quha}
 \ee
This relation would, indeed, render all of our observables
$\Lambda_j$ self-adjoint in ${\cal H}$ and, at the same time,
non-Hermitian and, in a way specified by definition (\ref{quha}),
``quasi-Hermitian'' in ${\cal F}$ \cite{Geyer,Dieudonne}.

This being said,
the feasibility of the NIP-based model-building strategy is still marred
by the fairly complicated nature of the description of the evolution
using the two independent generators $G(t)$ and $\Sigma(t)$.
This is one of the central questions and
challenges in the theory.
A simplification of the formalism is needed and sought in the
various methodically (rather than phenomenologically) motivated
restrictions of the admissible classes of eligible non-Hermitian
Hamiltonians.
In what follows we will analyze and describe, systematically,
the possibilities of
such a simplification.

\section{The first, dynamical-input strategy\label{hamb}}

%: $H_{(one)}(t)-$based picture old stationary

In any model-building scenario reflecting relation
(\ref{decoim}) which connects the three different ``Hamiltonians'' one
may start from an ``input'' knowledge of any one of them.
Still, the ``input information'' selection of the observable instantaneous
energy, viz., of the operator
 \be
 H_{(one)}^{(NIP)}(t)\neq 0
 \label{eone}
 \ee
looks most natural.
Such an option could be called ``dynamical'',
being most closely connected with the philosophy of Scholtz et al
\cite{Geyer} who treated all of their generalized
non-Hermitian quantum models as specified by their observables.

Once we restrict attention just to the observable Hamiltonian,
we have to deliver, first of all, a rigorous
proof of the reality of the energies.
Secondly, a consistent probabilistic interpretation of the model
requires
a
confirmation of the quasi-Hermiticity of the
observable Hamiltonian.
Thus, whenever the input information is encoded in operator
$H_{(one)}(t)$ (as well as in its conjugate form
$H^\dagger_{(one)}(t)$), our first task is to solve the
quasi-Hermiticity-constraint equation
 \be
 H^\dagger_{(one)}(t)\,\Theta_{(one)}(t)
 =\Theta_{(one)}(t)\,H_{(one)}(t)\,
 \label{tretiqh}
 \ee
for an unknown metric $\Theta_{(one)}(t)$.

The solution of such a linear algebraic problem
is non-unique but, conceptually,
straightforward. At any
time $t$, in a way indicated in \cite{NIP},
we may simply follow the notation convention of review
\cite{SIGMA} and initiate the search for
all (or at least for some) of the admissible
metric operators $\Theta_{(one)}(t)$
by solving the two
instantaneous Schr\"{o}dingerian eigenvalue problems
 \be
 H_{(one)}(t)\,|\psi^{(one)}(t)\kt=E_\psi(t)\,|\psi^{(one)}(t)\kt\,,
 \ \ \ \ \
 H^\dagger_{(one)}(t)\,|\psi^{(one)}(t)\kkt
 =E_\psi(t)\,|\psi^{(one)}(t)\kkt\,.
 \label{twoSE}
 \ee
In our notation, the symbol $\psi$ can be read either as an index
(numbering the elements of a complete set of states)
or, more traditionally,
as a letter which identifies a state in its
two different and complementary (i.e., single-ket and doubled-ket)
realizations.

In both of the equations in~(\ref{twoSE})
the energy eigenvalues
remain the same because they are, by assumption, observable (i.e., real),
discrete (because they have
to represent bound states \cite{Jones}) and
bounded from below (because the
system in question is assumed stable \cite{Carl}).
Nevertheless, due to the non-Hermiticity
$H_{(one)}(t)
\neq H^\dagger_{(one)}(t)$
of the Hamiltonian,
the respective two sets of the eigenvectors in (\ref{twoSE}) are
different.

In the context of physics
the knowledge of both of them is necessary
because both of them
contribute to the probabilistic predictions, i.e.,
to the matrix elements
 \be
 \bbr \psi(t)|\Lambda(t)|\psi(t)\kt
 \label{eme}
 \ee
in which
$\Lambda(t)$ denotes any observable of interest and in which
$t=t_f$ is a time of its measurement.
In the language of mathematics this means that
what is needed for a definition of a state is in fact an
elementary dyadic projector
 \be
 \pi_\psi(t)=|\psi^{(one)}(t)\kt\,
 \frac{1}{\bbr \psi^{(one)}(t)
 |\psi^{(one)}(t)\kt}\,\bbr \psi^{(one)}(t)|
 \label{epro}
 \ee
rather than just one of the two alternative
versions of the state vector.
This being clarified
we may
recall
their biorthogonality property \cite{Brody,SIGMAdva} and,
via a suitable rescaling, we may upgrade it to a
biorthonormality and bicompleteness,
 \be
 \bbr \psi^{(one)}(t)
 |\phi^{(one)}(t)\kt=\delta_{\psi \phi}\,,
 \ \ \ \
 \sum_\psi\,
 |\psi^{(one)}(t)\kt
 \bbr \psi^{(one)}(t)| = I\,.
 \label{onup}
 \ee
Formally, {\em all\,} of the metrics compatible with Eq.~(\ref{tretiqh})
can be then expressed in terms of the wave-function solutions of
the second, conjugate-operator equation in (\ref{twoSE}) \cite{SIGMAdva},
 \be
 \Theta_{(one)}(t)=
 \sum_{\psi}\,|\psi^{(one)}(t)\kkt\,
 \kappa_n^{(one)}(t)\,\bbr \psi^{(one)}(t)|\,.
 \label{onemet}
 \ee
It is easy to verify that in such a formula,
all of the parameters $\kappa_n^{(one)}$
are arbitrary. For the reasons as explained in \cite{Geyer},
they only have to be real and positive. Also,
for the sake of keeping the formalism
reasonably tractable (see a more explicit formulation of this
reason in \cite{NIP}), their recommended
choice will be time-independent, $\kappa_n^{(one)}(t)=\kappa_n^{(one)}(0)$.

The variability of the latter parameters can be interpreted either as
a formal kinematical freedom of the theory
(see, e.g., \cite{Lotor}) or, better,
as a manifestation of the above-mentioned incompleteness of the
dynamical-input information when restricted to
the single observable $H_{(one)}(t)$.
Indeed,
the formalism admits (and also, for the sake of completeness, requires)
an additional information about dynamics
simulated
by the choice of parameters $\kappa_n^{(one)}$.
More consequently and directly such an information could and should be, of course,
provided by the introduction of
additional observables
-- see
a more detailed discussion of
the suppression of the ambiguity in \cite{Geyer}.

\section{The second, Coriolis-choice strategy\label{Corio}}

The oldest formulation of
quantum mechanics, viz., the so called Heisenberg picture (HP, \cite{Messiah})
which appeared in June 1925 \cite{Styer}
can be characterized
as ``kinematical'' since
a strict time-independence of the wave functions is required, $\psi(t)=\psi(0)$.
In the Hermitian theory
one simply puts $G^{(HP)}(t)=0$ so that
relation~(\ref{decoim}) degenerates to
the identity $H^{(HP)}(t)=\Sigma^{(HP)}(t)$.
Just the above-discussed
observable-Hamiltonian dynamical input
is reobtained.

In a properly generalized non-Hermitian
NIP setup the situation is different \cite{NHeisenberg}.
A formulation of
strategy number two
becomes less
straightforward.
Although the ``kinematical''
design of models may still
start from the initial
specification
of the NIP Coriolis force at all of the relevant times $t$,
    \be
    \Sigma^{(NIP)}_{(two)}(t)\neq 0\,
 \label{giventwo}
    \ee
one has to
admit also  a
non-triviality of $G^{(NIP)}_{(two)}(t)\neq 0$
and of $H^{(NIP)}_{(two)}(t)\neq \Sigma^{(NIP)}_{(two)}(t)$
in~(\ref{decoim}).

Once we pick up
the kinematics (i.e.,
operator $\Sigma_{(two)}(t)$ and/or its conjugate
partner $\Sigma^\dagger_{(two)}(t)$),
we immediately imagine
that this opens the way to the
reconstruction of the unknown Dyson-mapping operator $\Omega_{(two)}(t)$.
For the purpose, indeed, it is sufficient
to recall its definition
(see the detailed introduction of this concept in \cite{SIGMA})
and to re-write it
in the following equivalent form
 \be
 {\rm i}\frac{d}{dt}\Omega_{(two)}(t)
 =\Omega_{(two)}(t)\,\Sigma_{(two)}(t)\,
 \label{prekdytwo}
 \ee
of the operator differential equation of the first order.
Its solution yields the Dyson-map operator at all times
from any preselected initial value at $t=0$.

For the sake of symmetry we may either conjugate the solution
or solve
the
conjugate
problem
 \be
 {\rm i}\frac{d}{dt}\Omega^\dagger_{(two)}(t)
 =-\Sigma^{\dagger}_{(two)}(t)\,\Omega^\dagger_{(two)}(t)\,.
 \label{rekdytwo}
 \ee
In the subsequent step we become able to define
the correct physical Hilbert-space metric as the product
of the two Dyson maps,
 \be
 \Theta_{(two)}(t)=\Omega^\dagger_{(two)}(t)\,\Omega_{(two)}(t)\,.
 \label{rekmetwo}
 \ee
The evaluation of this operator product
enables us to specify all of the eligible
Hamiltonians $H_{(two)}(t)$ as
(naturally, non-unique \cite{Geyer,SIGMAdva})
solutions of the Dieudonn\'{e}'s \cite{Dieudonne} quasi-Hermiticity
constraint
 \be
 H^\dagger_{(two)}(t)\,\Theta_{(two)}(t)
 =\Theta_{(two)}(t)\,H_{(two)}(t)=A_{(two)}(t)\,.
 \label{cojea}
 \ee
In a way inspired by the
non-Hermitian random-matrix theories \cite{Joshua}
we introduced here a new operator $A_{(two)}(t)$ encoding the
input information about dynamics
which is still
missing.

It is worth emphasizing that the latter
operator is almost arbitrary, restricted merely by
the requirement (\ref{cojea}) tractable as its
Hermiticity, $A_{(two)}(t)=A^\dagger_{(two)}(t)$.
This immediately yields the ultimate explicit
definitions
of both of the remaining unknown
components of the model,
 \be
 H_{(two)}(t)=\Theta^{-1}_{(two)}(t)\,A_{(two)}(t)\,,
 \ \ \ \ \ G_{(two)}(t)=H_{(two)}(t)-\Sigma_{(two)}(t)\,.
 \label{retwoall}
 \ee
The construction is completed.

\section{The third, state-evolution strategy\label{Gpic}}

% Schr\"{o}dinger-equation

Although the Erwin Schr\"{o}dinger's formulation of
quantum mechanics is not the oldest one \cite{Styer}, its extreme
conceptual as well as computational appeal and simplicity make
it a dominant
paradigm in textbooks \cite{Messiah}.
For this reason even Hynek B\'{\i}la,
one of my previous PhD
students
refused the terminology and philosophy
of my papers \cite{timedep,SIGMA} and,
even in the NIP regime, he
insisted on using
the dedicated name
``Hamiltonian'', strictly, just
for the
denotation of the
Schr\"{o}dinger's ``input physical information''
generator
of the evolution of wave functions \cite{PhD,Bila},
  \be
  G_{(three)}^{(NIP)}(t)\neq 0\,.
 \label{ethree}
 \ee
The  B\'{\i}la's
convention has later been accepted
by a number of other researchers \cite{ali,FringMou,PhDRT}.
They were influenced by the widely accepted
{\em stationary\,} non-Hermitian quantum theory of reviews
\cite{Geyer,ali,Carl} in which the time-independent
Schr\"{o}dinger's
operator $G(t)=G(0)$ coincides with
its stationary energy-representing partner $H(t)=H(0)$.
Obviously, this operator still had a
real spectrum and
carried a fully consistent physical meaning of an
observable.

Due to a rather naive straightforward transfer of
terminology to non-stationary scenarios the
key role has been allocated to $G(t)$ even when
$G(t) \neq H(t)$.
Incidentally, the change of the convention
appeared to have also several
positive aspects and consequences.
The main one was that after
a shift of attention
from the closed-system theory
to the open-system theory \cite{Nimrod}
or even beyond the domain of quantum physics \cite{Makris,Makrisb},
the loss of the
observability status of $G(t)$
(cf. its proof in Theorem~2
of review \cite{ali})
became irrelevant.
Thus,
in the study
of non-unitary,
open quantum systems, serendipitously, the B\'{\i}la's
terminology proved inspiring. A
number of interesting innovations of the traditional
mathematical concepts has been
revealed in this area: Cf., e.g., the
new use of the Lewis-Riesenfeld invariants as described in \cite{Maamache},
an innovative
introduction of a generalized
entropy in non-Hermitian systems in
\cite{eternal}, a reinterpretation of the concept of
${\cal PT}-$symmetry in \cite{PTFR} or, last but not least,
a new wave of interest in
non-linear theories, quantum (i.e.,
effective \cite{Christodoulides}) as well as
non-quantum \cite{Carlbook}.

Even after one returns back to the description of the closed
quantum systems, the initial selection of $G(t)$
need not destroy an internal consistency of the theory.
A detailed description of the related technicalities
may be found, e.g., in section 5.3.2 of paper \cite{universe}.
The point is that even the randomly emerging
complexifications of the spectrum of $G(t)$ may be kept compatible
with the
unitarity of evolution of the underlying closed quantum system.
Via
a few schematic non-stationary toy models
this was illustrated in
\cite{2by2,3by3}.

A model-independent methodical support
of the $G(t)-$based approach can be
based on our preceding considerations.
First of all, we have to return to the
concept of a biorthonormal and bicomplete basis.
Our assumption of the knowledge of operator (\ref{ethree})
at all times
implies that
it is now sufficient to know
the pure-state-representing
projector (\ref{epro}) just
at $t=0$,
 \be
 \pi_\psi(0)=|\psi^{(three)}(0)\kt\,
 \frac{1}{\bbr \psi^{(three)}(0)
 |\psi^{(three)}(0)\kt}\,\bbr \psi^{(three)}(0)|\,.
 \label{epro3}
 \ee
From the point of view of physics this means that
the theory admits the preparation of a more or less
arbitrary initial state of the quantum system in question.

In the next step
we may now recall the availability of $G(t)$
and solve the two evolution-equation analogues
 \be
 {\rm i}\,\frac{d}{dt}\,|\psi^{(three)}(t)\kt
 =
 G_{(three)}(t)\,|\psi^{(three)}(t)\kt
 \,
 \label{Cauchy}
 \ee
and
 \be
 {\rm i}\,\frac{d}{dt}\,|\psi^{(three)}(t)\kkt=
 G^\dagger_{(three)}(t)\,|\psi^{(three)}(t)\kkt\,
 \label{Cauchybe}
 \ee
of the two Schr\"{o}dingerian eigenvalue problems (\ref{twoSE})
(see also the details in \cite{timedep}).
Next, via an appropriate rescaling of the
initial-value vectors
we may
achieve their bi-orthonormality and bicompleteness,
 \be
 \bbr \psi^{(three)}(0)
 |\phi^{(three)}(0)\kt=\delta_{\psi \phi}\,,
 \ \ \ \
 \sum_\psi\,
 |\psi^{(three)}(0)\kt
 \bbr \psi^{(three)}(0)| = I\,.
 \label{thrup0}
 \ee
Finally, recalling the theory \cite{NIP}
we may extend the validity
of these postulates to all times $t$,
 \be
 \bbr \psi^{(three)}(t)
 |\phi^{(three)}(t)\kt=\delta_{\psi \phi}\,,
 \ \ \ \
 \sum_\psi\,
 |\psi^{(three)}(t)\kt
 \bbr \psi^{(three)}(t)| = I\,.
 \label{thrup}
 \ee
Partial methodical parallels with the dynamical-input strategy
emerge: Whenever our initial-time choice
of the biorthonormalized and bicomplete
basis of Eq.~(\ref{thrup0})
is made compatible with
the input-information form of one of the observables,
(i.e., say, of the energy operator) at $t=0$,
 \be
 H_{(three)}(0)=
 \sum_\psi\,
 |\psi^{(three)}(0)\kt\,E_\psi^{(three)}(0)\,
 \bbr \psi^{(three)}(0)|\,,
 \label{ethrup0}
 \ee
we may immediately reconstruct the same
operator
at all times $t>0$. Naturally,
also the construction of the metric acquires the explicit form
 \be
 \Theta_{(three)}(t)=
 \sum_{\psi}\,|\psi^{(three)}(t)\kkt\,
 \kappa_n^{(three)}(t)\,\bbr \psi^{(three)}(t)|\,.
 \label{thremet}
 \ee
At this stage of development of the theory it is useful to notice that
many of its applications (cf., e.g., \cite{Bishop})
are just considered in a finite-dimensional Hilbert space. Then,
many operators of interest (i.e., $N$ by $N$ matrices with $N < \infty$)
may happen to
form a representation of a suitable Lie algebra. This does not only render the
constructions feasible but it also enables us to factorize the metric
into a product of Dyson maps,
 \be
 \Theta_{(three)}(t)=\Omega^\dagger_{(three)}(t)\,\Omega_{(three)}(t)\,
 \label{threkme}\,.
  \ee
The latter formula may be compared with Eq.~(\ref{rekmetwo}) where the
construction proceeded from $\Omega$ to $\Theta$, i.e., in
the opposite direction.
In other words, the Dyson map may formally be written as the sum
 \be
 \Omega_{(three)}(t)=\sum_\psi\,|\psi_{(three)}(t)\pkt
 \sqrt{\kappa_n^{(three)}(0)}\,\,\bbr \psi_{(three)}(t)|
 \label{forDy}
 \ee
where
the new auxiliary basis $\{|\psi_{(three)}(t)\pkt\}$
may be chosen orthonormal.

On this level of reconstruction one is already able to define the Coriolis force,
 \be
 \Sigma_{(three)}(t)=\frac{\rm i}{\Omega_{(three)}(t)}\,\dot{\Omega}_{(three)}(t)
 \ee
where the dot represents the differentiation with respect to time. Now, the last step
yielding, finally, the observable Hamiltonian and its decomposition
 \be
 H_{(three)}(t)=G_{(three)}(t)+\Sigma_{(three)}(t)
 \ee
is already trivial.

\section{Summary\label{progresivni}}

It is well known that the price to be paid for the generality and
flexibility of the
NIP formulation of non-stationary quantum theory
in both of its quantum-mechanical and field-theoretical versions
is not too low. Only too many
evolution equations have to be solved. In our present paper
we managed to show that a systematic subdivision of the
related model-building strategies could simplify the picture thoroughly.
The core of our message lies in the observation
that the most natural interpretation of all of the eligible
NIP-based mathematical constructions
of quantum models
should be based on a
clear separation of the consistent implementation alternatives.

We have shown that an explicit guide to the choice out of the menu
has to be correlated with
a context-dependent dominance of one of the
operators $H(t)$, $\Sigma(t)$ or $G(t)$.
We argued that such an identification of dominance
leads directly to
the three different NIP-implementation recipes.
They may be characterized by their
specific respective mathematical merits
as well as by a natural subdivision and classification of the
related phenomenological intentions.
Thus, what we described are the three mutually complementary
forms
of the implementation of the abstract NIP quantum theory
in the situations where the set of the underlying
unitary (i.e., closed) quantum systems
can be subdivided according to the
more detailed practical criteria.

The resulting
construction
process seems useful, enhancing the
tractability of the systems living in
a non-stationary dynamical regime in which the use of
the hiddenly Hermitian representations of observables
might throw new light also on the physical
interpretation of the various important open questions,
say, in cosmology \cite{axioms}.
In all of these contexts,
a clear separation of the input information
about the system from the resulting predictions
seems to play, in non-stationary systems,
a more important role than in their stationary predecessors
because the increase of the complexity of mathematics is
enormous. The traditional guidance by the stationary constructions using
analogies with the techniques of linear algebra becomes,
in the NIP framework, replaced by the necessity of solving
complicated operator evolution equations.
We believe that such a challenge has to be accepted.
As a reward, indeed,
the NIP formalism may be expected to open the
new ways of description of multiple deeply non-stationary
phenomena.

\newpage

\section*{Appendix A: A few remarks on applications}

In full accord with the traditional textbooks on quantum mechanics
\cite{Messiah} the observables  have to be
self-adjoint in
a suitable Hilbert space.
Nevertheless, in the  non-stationary quantum NIP approach such a
``correct'' space (say, ${\cal L}_{(o\!f\ textbooks)}$) is
``hidden'' (incidentally, one of the traces of its
hidden existence may be seen in the emergence of the
auxiliary basis $\{|\psi_{(three)}(t)\pkt\}$ in formula (\ref{forDy})).
Its role is
symbolic, reduced to the mere hypothetical reference
to the conventional probabilistic-interpretation contents of the model.

In
\cite{Entropy}
we paid attention to the applicability and application of the NIP
evolution equations to several versions of the so called
``wrong-sign'' quartic oscillator, with the choice inspired by a few
older studies \cite{Jones,BG,FT}.
We emphasized that it is absolutely necessary to separate,
very carefully, the study of the stationary systems from the study
of their non-stationary analogues.

One of the reasons has already
been known to Mostafazadeh who considered, on p. 1271 of his
comprehensive 2010 review \cite{ali}, a time-dependent and
non-Hermitian ``Schr\"{o}dinger-equation Hamiltonian'' $H^{(SEH)}(t)$
(abbreviated, in our present notation, as $G(t)$) and proved,
in Theorem Nr. 2 of {\it loc. cit.}, that ``if the time evolution of
the system \ldots is unitary and [if] $H^{(SEH)}(t)$ is an
observable for all $t\in (0,T)$, then the metric operator operator
[$\Theta$] defining ${\cal H}_{(physical)}\equiv {\cal H}$ does not
depend on time''.

In our subsequent commentaries (cf., e.g., \cite{NIP}) we fully
agreed with the latter, strictly mathematical result. At the same
time we emphasized that in the general quantum-mechanical NIP
theoretical framework there exists absolutely no reason for the
purely formalistic, fully redundant and entirely unfounded
requirement of the observability of the
Schr\"{o}dinger-equation-evolution generator $G(t)
\equiv\,H^{(SEH)}(t)$.
In \cite{Entropy} we complemented such a statement by its
``wrong-sign'' quartic-oscillator illustration.

In our present,
methodically oriented
continuation of the latter paper we just gave this statement
a more explicit and more systematic form. We emphasized that
both the Schr\"{o}dinger-equation generators $G(t)$ and their
Heisenberg-equation analogues $\Sigma(t)$ are just auxiliary concepts.
They enter the respective NIP
evolution equations but,
in contrast to the widespread belief, their
spectra are, in general,
complex. For this reason, only their sum $H(t) = G(t) + \Sigma(t)$
is, from the point of view of physics, relevant,
retaining the standard physical meaning of the
instantaneous energy of the unitary quantum system in question.
The postulate of its choice makes the construction strategy number one
phenomenologically preferable.

From the point of view of mathematics, the situation is different because
the experimental predictions are only given by
the evaluation of the mean values of the form of overlap Eq.~(\ref{eme}).
In the NIP framework, for this reason, one must evaluate,
at the time of measurement $t=t_f$,
{\em both\,} the (elementary) projector $\pi_\psi(t)$ of Eq.~(\ref{epro})
{\em and\,} the (complicated) operator $\Lambda(t_f)$.
For this reason, the evaluation of the predictions (to be tested by a
hypothetical experiment) requires not only the (usually, emphasized)
solution of the two
Schr\"{o}dinger-type evolution equations (viz,
Eqs.~(\ref{Cauchy}) and (\ref{Cauchybe})) for vectors
but also a parallel
solution of the
Heisenberg-type evolution equation for $\Lambda(t)$
as given, e.g., by equation Nr. 34 in \cite{Entropy}.

In practice, the solvability
of the two
Schr\"{o}dinger-type evolution equations
(characterizing the state of the system
via dyadic $\pi_\psi(t)$)
is a comparatively easy part of the task because
such a projector is defined just in terms
of the
two formally independent
{\em state vectors\,} $|\psi(t)\kt$ and  $|\psi(t)\kkt$.
In comparison, the necessary simultaneous construction of
$\Lambda(t)$ is in fact
by far the most difficult part of the
task.
Indeed, even the most simple, conservative-observable version of
the related operator evolution equation
having the familiar Heisenberg's form
 \be
 {\rm i\,}\frac{\partial}{\partial t} \,{Q}_{}(t)=
  Q(t)\,\Sigma(t) -\Sigma_{}(t)\,Q(t)
 %+K(t)\,,
% \ \ \ \ \ K(t)=\Omega_{}^{(-1)}(t)\,
% {\rm i\,}\dot{\mathfrak{q}}_{(SP)}(t)\,\Omega_{}(t)\,.
 \label{beda}
 \ee
has to specify the solution which is an {\em operator}.

The latter observation could even be used as an argument in favor of
the preference of the kinematic, Coriolis-based strategy number two.
In it, indeed, one can fine-tune the operator $\Sigma_{}(t)$ to make it
as user-friendly as possible. Another, analogous supportive argument
could be also found in the Fring's and Tenney's paper \cite{FT}
in which the authors used a version of such a strategy for
successful and explicit, non-numerical
toy-model constructions. In fact, their procedure
only deviated from
the present, Coriolis-based one in a
re-interpretation of Eq.~(\ref{prekdytwo}). In their case
a decisive role of a success-yielding input information has been
played by a trial-and-error
ansatz for
$\Omega_{(two)}(t)$ (cf. Eq. Nr. (2.3) in {\it loc. cit.})
rather than by our above-recommended ansatz for
$\Sigma_{(two)}(t)$.

In our comment \cite{Entropy} on the latter construction
we pointed out that in the case of the ``wrong-sign''
anharmonic toy models
several equally compact algebraic results may be also obtained using
directly the
energy-based strategy number one.
The key role in the description of dynamics has been played
there by the
``observable-Hamiltonian'' operator sum
$G(t)+\Sigma(t)=H(t)$.
This enabled us to simplify the
calculations, especially when the operators of observables
remained time-independent, $\Lambda_j(t)=\Lambda_j(0)$.

During the development of the NIP theory after 2008
it became known that the condition of stationarity
$H\neq H(t)$ appears to be not so easy to relax
(see, e.g., \cite{PLB,Faria,Gong} or Theorem Nr. 2 in review \cite{ali}).
In fact, the emerging difficulties were of two types.
The main ones were abstract and
concerned the consistency of the theory. The
equally important obstacles emerged, in the context
of the applicability of the formalism,
as a consequence of the
operator nature of the NIP evolution equations.

The answers to the abstract conceptual questions
were not too difficult and their majority may be found
provided in the early paper \cite{timedep} and in review \cite{SIGMA}.
In contrast, the emergence of the NIP-related technical challenges
resembled the opening of a
Pandora's box: In place of the mere construction
of the metric
in stationary models,
it appeared necessary to solve several incomparably more complicated
evolution equations \cite{Bila}.

The essence of the message was that
one merely has to leave the over-restrictive Schr\"{o}dinger-picture framework
and that a consistent formulation of the non-stationary non-Hermitian theory
requires just a transition to a non-Hermitian analogue
of intermediate picture (NIP).
A consistent and unitary non-stationary {\it alias\,}
NIP quantum theory has been formulated in which only the
observability of the ``observable non-stationary quasi-Hermitian
Hamiltonian'' $H(t)=G(t)+\Sigma(t)$ is in fact required
and needed.
Unfortunately, as long as all of the three operator components in
the latter sum can be called ``Hamiltonians'', a series of
misunderstandings followed.

In 2010, in particular, one could still
read, in the mathematically rigorous review \cite{ali},
that in the time-dependent cases, ``insisting on observability of
the Hamiltonian operator'' would be inconsistent (see Theorem 2 in
{\it loc. cit.}).
The source of such a misunderstandings can be traced back to the
terminology. Indeed, in a
way paralleling the B\'{\i}la's 2009 proposal \cite{Bila} and in a
way used, later, also by Fring et al
\cite{FringMou,FT,infinite}, the author of review \cite{ali} did
not in fact have in mind the ``observable non-stationary
quasi-Hermitian Hamiltonian'' $H(t)=G(t)+\Sigma(t)$ but
rather just another, purely auxiliary operator representing, in our
present notation, the time-dependent Schr\"{o}dinger-equation
generator {\it alias\,} ``unobservable Hamiltonian''
$G(t)$.
The related dedicated discussions resolved the paradox and
helped to clarify the puzzle.

\end{document}